\documentclass[journal,twocolumn]{IEEEtran}
\usepackage{stmaryrd}
\usepackage{amsfonts}
\usepackage{amsmath}
\usepackage{amssymb}
\usepackage{cite}
\usepackage{graphicx}
\usepackage{subfigure}
\usepackage{amsthm}
\usepackage{algorithmic}
\usepackage{algorithm}
\usepackage{stfloats}

\newtheorem{Theorem}{Theorem}

\newtheorem{Lemma}[Theorem]{Lemma}
\newtheorem{Corollary}[Theorem]{Corollary}

\newcommand{\nchoosek}[2]{\binom{#1}{#2}}

\newcommand{\newQED}{\IEEEQED \vspace{0.1in}}

\begin{document}
%

\title{A Universal Scheme for Transforming Binary Algorithms to Generate Random Bits from Loaded Dice}

\author{Hongchao~Zhou,
        and~Jehoshua~Bruck,~\IEEEmembership{Fellow,~IEEE}
\thanks{Hongchao~Zhou and Jehoshua~Bruck are with the Department
of Electrical Engineering, California Institute of Technology, Pasadena,
CA 91125, USA, e-mail: hzhou@caltech.edu; bruck@caltech.edu.}
\thanks{This work was supported in part by the NSF Expeditions in Computing
Program under grant CCF-0832824.}
}


\maketitle%
\begin{abstract}
In this paper, we present
a universal scheme for transforming an arbitrary algorithm for biased $2$-face coins to generate random bits from  the general
source of an $m$-sided die, hence enabling the application of existing algorithms to general sources.
In addition, we study approaches of efficiently generating a prescribed number of random bits from an arbitrary biased coin. This contrasts with
most existing works, which typically assume that the number of coin tosses is fixed, and they generate a variable number of random bits.
\end{abstract}


\IEEEpeerreviewmaketitle

\section{Introduction}

\IEEEPARstart{I}{n} this paper, we study the problem of random number generation from i.i.d. sources, which is the most fundamental and important source model. Many real sources can be well approximated by this model, and the algorithms developed based on this model can be further generalized in generating random bits from more sophisticated models, like Markov chains \cite{Zhou2012Markov}, or more generally, approximately stationary ergodic processes \cite{Zhou_variablextractor}.

The problem of random number generation dates back to von Neumann \cite{Neumann1951} in 1951 who considered the problem of simulating an unbiased coin by using a biased coin with unknown probability. He observed that when one focuses on a pair of coin tosses, the events $\textrm{HT}$ and $\textrm{TH}$ have the same probability ($\textrm{H}$ is for `head' and $\textrm{T}$ is for `tail'); hence, $\textrm{HT}$ produces the output symbol $1$ and $\textrm{TH}$ produces the output symbol $0$. The other two possible events, namely, $\textrm{HH}$ and $\textrm{TT}$, are ignored, namely, they do not produce any output symbols. More efficient algorithms for generating random bits from a biased coin were proposed by Hoeffding and Simons \cite{Hoeffding1970}, Elias \cite{Elias1972}, Stout and Warren \cite{Stout1984} and Peres \cite{Peres1992}. Elias \cite{Elias1972} was the first to devise an optimal procedure in terms of the information efficiency, namely, the expected number of unbiased random bits generated per coin toss is asymptotically equal to the entropy of the biased coin. In addition, Knuth and Yao \cite{Knuth1976} presented a simple procedure for generating sequences with arbitrary probability distributions from an unbiased coin (the probability of $\textrm{H}$ and $\textrm{T}$ is $\frac{1}{2}$). Han and Hoshi \cite{Han1997} generalized this approach and considered the case where the given coin has an arbitrary known bias.

In this paper, we consider the problem of generating random bits from a loaded die as a natural generalization of generating random bits from a biased coin. There is some related work: In \cite{Dijkstra1990}, Dijkstra considered the opposite question and showed how to use a biased coin to simulate a fair die. In \cite{Juels2000}, Juels et al. studied the problem of simulating random bits from loaded dice, and their algorithm can be treated as the national generalization of Elias's algorithm. However, for a number of known beautiful algorithms, like Peres's algorithm \cite{Peres1992}, we still do not know how to generalize them for larger alphabets (loaded dice).

In addition, we notice that most existing works for biased coins take a fixed number of coin tosses as the input and they generate a variable number of random bits. In some occasions, the opposite question seems more reasonable and useful: given a biased coin, how to generate a prescribed number of random bits with as a few as possible coin tosses? Hence, we want to create
a function $f$ that maps the sequences in a dictionary $\mathcal{D}$, whose lengthes may be different, to binary sequences of the same length.
This dictionary $\mathcal{D}$ is complete and prefix-free. That means for any infinite sequence, it has exactly one prefix in the dictionary. To generate random bits, we read symbols from the source until the current input sequence matches one in the dictionary.

For completeness, in this paper, we first present some of the existing algorithms that generate random bits from an arbitrary biased coin in Section \ref{section_coin_existing}, including the von Neumann Scheme, Elias algorithm and Peres algorithm. Then in Section \ref{section_coin_generalization}, we present a universal scheme for transforming an arbitrary algorithm for $2$-faced coins to generate random bits from the general
source of an $m$-sided die, hence enabling the application of existing algorithms to general sources.
In Section \ref{section_coin_kbits}, we study approaches of efficiently generating a required number of random bits from an arbitrary biased coin
and achieving the information-theoretic upper bound on efficiency. Finally, we provide the concluding remarks in
Section \ref{section_coin_conclusion}.

\section{Existing Algorithms for Biased Coins}
\label{section_coin_existing}

\subsection{Von Neumann Scheme}

In 1951, von Neumann \cite{Neumann1951} considered the problem of random number generation from biased coins and described
a simple procedure for generating an independent unbiased binary sequence $z_1z_2...$ from an input sequence $X=x_1x_2...$.
His original procedure is described as follows:  For an input sequence, we divide all the bits into pairs $x_1x_2, x_3x_4, ...$ and apply the following mapping to each pair
$$\textrm{HT}\rightarrow 1,\quad \textrm{TH}\rightarrow 0, \quad \textrm{HH} \rightarrow \phi, \quad \textrm{TT}\rightarrow \phi,$$
where $\phi$ denotes the empty sequence. By concatenating the outputs of all the pairs, we can get a binary sequence, which is independent and unbiased.
The von Neumann scheme is computationally (very) fast, however, its information efficiency is far from being optimal. Here, the information efficiency
is defined by the expected number of random bits generated per input symbol. Let $p_1, p_2$ with $p_1+p_2=1$ be the probabilities of getting H and T, then
the probability for a pair of input bits to generate
one output bit (not a $\phi$) is $2p_1p_2$, hence the information efficiency is $\frac{2p_1p_2}{2}=p_1p_2$, which
is $\frac{1}{4}$ at $p_1=p_2=\frac{1}{2}$ and less elsewhere.

\subsection{Elias Algorithm}

In 1972, Elias \cite{Elias1972} proposed an optimal (in terms of information efficiency) algorithm
as a generalization of the von Neumann scheme.

Elias's method is based on the following idea: The possible $2^n$ binary input sequences of length $n$ can be partitioned into classes
such that all the sequences in the same class have the same number of H's and T's. Note that for every class, the members of the class have the same probability to be generated. For example, let $n=4$, we can divide the possible $2^n=16$
input sequences into $5$ classes:
\begin{align*}
  &S_0= \{\textrm{HHHH}\},\\
  &S_1= \{\textrm{HHHT}, \textrm{HHTH}, \textrm{HTHH}, \textrm{THHH}\},\\
  &S_2= \{\textrm{HHTT}, \textrm{HTHT}, \textrm{HTTH}, \textrm{THHT}, \textrm{THTH}, \textrm{TTHH}\},\\
  &S_3=\{\textrm{HTTT}, \textrm{THTT}, \textrm{TTHT}, \textrm{TTTH}\},\\
  &S_4=\{\textrm{TTTT}\}.
\end{align*}

Now, our goal is to assign a string of bits (the output) to each possible input sequence, such that any two possible output sequences $Y$ and $Y'$ with the same length (say $k$), have the same probability to be generated, which is $\frac{c_k}{2^k}$ for some $0\leq c_k\leq 1$. The idea is that for any given class we partition the members of the class to sets of sizes that are a power of 2, for a set with $2^i$ members (for some $i$) we assign binary strings of length $i$. Note that when the class size is odd we have to exclude one member of this class. We now demonstrate the idea by continuing
the example above.

In the example above, we cannot assign any bits to the sequence in $S_0$, so if the input sequence is HHHH,
the output sequence should be $\phi$ (denoting the empty sequence). There are $4$ sequences in $S_1$ and we assign the binary strings as follows:
$$\textrm{HHHT}\rightarrow 00,\quad \textrm{HHTH}\rightarrow 01,$$
$$\textrm{HTHH}\rightarrow 10,\quad \textrm{THHH}\rightarrow 11.$$
Similarly, for $S_2$, there are $6$ sequences that can be divided into a set of $4$
and a set of $2$:
$$\textrm{HHTT}\rightarrow 00, \quad  \textrm{HTHT}\rightarrow 01,$$
$$\textrm{HTTH}\rightarrow 10, \quad  \textrm{THHT}\rightarrow 11,$$
$$\textrm{THTH}\rightarrow 0, \quad  \textrm{TTHH}\rightarrow 1.$$

In general, for a class with $W$ members that were not assigned yet, assign $2^j$ possible output binary sequences of length $j$ to $2^j$ distinct unassigned members, where
$2^j\leq W<2^{j+1}$.  Repeat the procedure above for the rest of the members
that were not assigned. When a class has an odd number of members, there will be one and only one member assigned to $\phi$.

Given a binary input sequence $X$ of length $n$, using the method above, the output sequence can be written as a function of $X$, denoted by $\Psi_{E}(X)$, called the Elias function.
In \cite{Ryabko2000}, Ryabko and Matchikina showed that the Elias function of an input sequence
of length $n$ (that is generated by a biased coin with two faces) is computable in $O(n \log^3 n \log\log (n))$ time.

\subsection{Peres Algorithm}

In 1992, Peres \cite{Peres1992} demonstrated that iterating the original von Neumann scheme
on the discarded information can asymptotically achieve optimal information efficiency.
Let us define the function related to the von Neumann scheme as $\Psi_1: \{\textrm{H},\textrm{T}\}^*\rightarrow
\{0,1\}^*$. Then the iterated procedures $\Psi_v$ with $v\geq 2$ are defined inductively.
Given an input sequence $x_1x_2...x_{2m}$, let $i_1<i_2<...<i_k$ denote all the integers $i\leq m$ for which $x_{2i}=x_{2i-1}$, then $\Psi_v$ is defined as
\begin{eqnarray*}
 && \Psi_v(x_1,x_2,...,x_{2m}) \\
  &=& \Psi_1(x_1,x_2,...,x_{2m})*\Psi_{v-1}(x_1\oplus x_2, ..., x_{2m-1}\oplus x_{2m}) \\
   && *\Psi_{v-1}(x_{2i_1}, ..., x_{2i_k}).
\end{eqnarray*}

Note that on the righthand side of the equation above, the first term corresponds to the random bits
generated with the von Neumann scheme, the second and third terms relate
to the symmetric information discarded by the von Neumann scheme. For example, when the input sequence is $X=\textrm{HHTHTT}$, the output sequence based on the von Neumann scheme
is
$$\Psi_1(\textrm{HHTHTT})=0.$$
But based on the Peres scheme, we have the output sequence
$$\Psi_v(\textrm{HHTHTT})=\Psi_1(\textrm{HHTHTT})*\Psi_{v-1}(\textrm{THT})*\Psi_{v-1}(\textrm{HT}),$$
which is $001$, longer than that generated by the von Neumann scheme.

Finally, we can define $\Psi_v$ for sequences of odd length by
$$\Psi_v(x_1,x_2,...,x_{2m+1})=\Psi_v(x_1,x_2,...,x_{2m}).$$

Surprisingly, this simple iterative procedure achieves the optimal information efficiency asymptotically.
The computational complexity and memory requirements of this scheme are substantially smaller than those of the Elias scheme. However,
the generalization of this scheme to the case of an $m$-sided die with $m>2$ is still unknown.

\subsection{Properties}

Let us denote $\Psi: \{\textrm{H},\textrm{T}\}^n \rightarrow \{0,1\}^*$ as a scheme that generates independent unbiased sequences from any biased coins (with unknown probabilities).
Such $\Psi$ can be the von Neumann scheme, the Elias scheme, the Peres scheme, or any other scheme. Let $X$ be a sequence of biased coin tosses of length $n$, then a property of $\Psi$ is that for any $Y\in \{0,1\}^*$ and $Y' \in \{0,1\}^*$ with $|Y|=|Y'|$, we have
$$P[\Psi(X)=Y]=P[\Psi(X)=Y'],$$
i.e., two output sequences of equal length have equal probability.

This observation leads to the following property for $\Psi$. It says that given the numbers of H's and T's,
the number of sequences yielding a binary sequence $Y$ equals the number of sequences yielding $Y'$ when $Y$ and $Y'$ have the same length.
It further implies that given the condition of knowing the number of H's and T's in the input sequence, the output sequence
of $\Psi$ is still independent and unbiased. This property is due to the
linear independence of probability functions of the sequences with different numbers
of H's and T's.

\begin{Lemma} \emph{\cite{Zhou2012Markov}}\label{lemma_coin}
Let $S_{k_1,k_2}$ be the subset of $\{\textrm{H},\textrm{T}\}^n$ consisting of all sequences with $k_1$ appearances of H and $k_2$ appearances of T such that
$k_1+k_2=n$. Let $B_Y$ denote the set $\{X|\Psi(X)=Y\}$. Then for any $Y\in \{0,1\}^*$ and $Y' \in \{0,1\}^*$ with $|Y|=|Y'|$, we have $$|S_{k_1,k_2}\bigcap B_Y|=|S_{k_1,k_2}\bigcap B_{Y'}|.$$
\end{Lemma}
\vspace{0.005in}

\section{Generalization for Loaded Dice}
\label{section_coin_generalization}

In this section, we propose a universal scheme for generalizing all the existing algorithms for biased coins such that
they can deal with loaded dice with more than two sides. There is some related work: In \cite{Dijkstra1990}, Dijkstra considered the opposite question and showed how to use a biased coin
to simulate a fair die. In \cite{Juels2000}, Juels et al. studied the problem of simulating random bits from loaded dice, and their algorithm can be treated as the generalization of Elias's algorithm. However, for a number of known beautiful algorithms, like Peres's algorithm, we still do not know how to generalize them for larger alphabets (loaded dice). We propose a universal scheme that is able to generalize all the existing algorithms, including Elias's algorithm and Peres's algorithm. Compared to the other generalizations, this scheme is universal and easier to implement, and it preserves the optimality of the original algorithm on information efficiency. The brief idea of this scheme is that given a loaded die, we can convert it into multiple binary sources and apply existing algorithms to these binary sources separately. This idea seems natural, but not obvious.

\subsection{An Example}
Let us start from a simple example: Assume we want to generate random bits from a sequence $X=012112210$, which is produced by a
$3$-sided die. Now, we write each symbol (die roll) into a binary representation of length two (H for 1 and T for 0), so
$$0\rightarrow \textrm{TT}, 1\rightarrow \textrm{TH},  2\rightarrow \textrm{HT}.$$

Hence, $X$ can be represented as
$$\textrm{TT,TH,HT,TH,TH,HT,HT,TH,TT}.$$

Only collecting the first bits of all the symbols yields an independent binary sequence $$X_{\phi}=\textrm{TTHTTHHTT}.$$
Collecting the second bits following T, we get another independent binary sequence $$X_\textrm{T}=\textrm{THHHHT}.$$
Note that although both $X_{\phi}$ and $X_\textrm{T}$ are independent sequences individually, $X_\phi$ and $X_\textrm{T}$ are correlated with each other, since
the length of $X_\textrm{T}$ is determined by the content of $X_{\phi}$.

Let $\Psi$
be any function that generates random bits from a fixed number of coin tosses, such as Elias's algorithm and Peres's algorithm.
We see that both $\Psi(X_\phi)$ and $\Psi(X_\textrm{T})$ are sequences of random bits. But we do not know whether $\Psi(X_\phi)$
and $\Psi(X_\textrm{T})$ are independent of each other since $X_\phi$ and $X_\textrm{T}$ are correlated.
One of our main contributions is to
show that concatenating them together, i.e., $$\Psi(X_\phi)+\Psi(X_\textrm{T})$$
still yields a sequence of random bits.

\subsection{A Universal Scheme}
\begin{figure}[!t]
\centering
\includegraphics[width=2.4in]{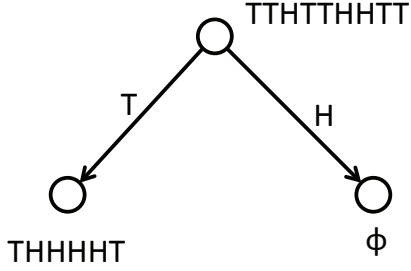}
\caption{An instance of binarization tree.}
\label{fig_prefixtree}
\end{figure}

Generally, given a sequence of symbols generated from an $m$-sided die, written as
$$X=x_1x_2...x_n\in \{0,1,...,m-1\}^n$$
with the number of states (sides) $m>2$, we want to convert it into a group of binary sequences. To do this, we create a binary tree, called a binarization tree, in which
each node is labeled with a binary sequence of H and T. See Fig.~\ref{fig_prefixtree} as an instance of binarization tree for the above example.
Given the binary representations of $x_i$ for all $1\leq i\leq n$,
the path of each node in the tree indicates a prefix, and the binary sequence labeled at this node
consists of all the bits (H or T) following the prefix in the binary representations of $x_1, x_2, ..., x_n$ (if it exists).

Given the number of sides $m$ of a loaded die, the depth of the binarization tree is $b=\lceil\log_2 m \rceil-1$. At the beginning, the binarization tree is a complete binary tree of depth $b$ in which each node is labeled with an empty string, then we process all the input symbols $x_1, x_2, ..., x_n$ one by one.
For the $i$th symbol, namely $x_i$, its binary representation is of length $b+1$. We add its first bit to the root node. If this bit is T, we add its
second bit to the left child, otherwise we add its second bit to the right child ... repeating this process until all the $b+1$ bits of $x_i$ are added along a path in the tree.
Finally, we can get the binarization tree of $X$ by processing all the symbols in $X$, i.e., $x_1, x_2, ..., x_n$.

\begin{Lemma} \label{lemma3_0}
Given the binarization tree of a sequence $X\in \{0,1,...,m-1\}^n$, we can reconstruct $X$ uniquely.
\end{Lemma}

\IEEEproof
The construction of $X$ from its binarization tree can be described as follows: At first, we read the first bit (H or T) from the root (once we read a bit, we remove it from the current sequence). If it is T, we read the first bit of its left child; if it is H, we read the first bit of its right child ... finally we reach a leaf, whose path indicates the binary representation of $x_1$. Repeating this procedure, we can continue to obtain $x_2, x_3, ..., x_n$.
\hfill\newQED

Let $\Upsilon_b$ denote the set consisting of all the binary sequences of length at most $b$, i.e.,
$$\Upsilon_b=\{\textrm{$\phi$, T, H, TT, TH, HT, HH, ..., HHH...HH}\}.$$
Given $X\in\{0,1,...,m-1\}^n$, let $X_\gamma$ denote the binary sequence labeled on a node corresponding to a prefix $\gamma$ in the binarization tree, then
we get a group of binary sequences
$$X_\phi, X_\textrm{T}, X_\textrm{H}, X_\textrm{TT}, X_\textrm{TH}, X_\textrm{HT}, X_\textrm{HH}, ...$$
For any function $\Psi$ that generates random bits from a fixed number of coin tosses, we can generate random bits from $X$ by calculating
$$\Psi(X_\phi)+\Psi(X_\textrm{T})+\Psi(X_\textrm{H})+\Psi(X_\textrm{TT})+\Psi(X_\textrm{TH})+...,$$
where $A+B$ is the concatenation of $A$ and $B$.
We call this method as the generalized scheme of $\Psi$.

We show that the generalized scheme works for any binary algorithm $\Psi$ such that
it can generate random bits from an arbitrary $m$-sided die.

\begin{Theorem} \label{theorem3} Let $\Psi$ be any function that generates random bits from a fixed number of coin tosses.
Given a sequence $X\in \{0,1,...,m-1\}^n$ with $m\geq 2$ generated from an $m$-sided die, the generalized scheme of $\Psi$ generates an independent and unbiased sequence.
\end{Theorem}

The proof of this theorem will be given in the next subsection.

\subsection{Proof of Theorem \ref{theorem3}}

\begin{Lemma} \label{Stream_lemma3_1}
Let $\{X_{\gamma}\}$ with $\gamma\in \Upsilon_b$ be the binary sequences labeled on the binarization tree of $X\in \{0,1,...,m-1\}^n$ as defined above.
Assume $X_\gamma'$ is a permutation of $X_\gamma$ for all $\gamma\in \Upsilon_b$, then there exists exactly one sequence $X'\in \{0,1,...,m-1\}^n$ such that it yields a binarization tree that labels $\{X_{\gamma}'\}$ with $\gamma\in \Upsilon_b$.
\end{Lemma}

\IEEEproof
Based on $\{X_{\gamma}'\}$ with $\gamma\in \Upsilon_b$, we can construct the corresponding binarization tree and then create the sequence $X'$ in the following way (if it exists).
At first, we read the first bit (H or T) from the root (once we read a bit, we remove it from the current sequence). If it is T, we read the first bit of its left child; if it is H, we read the first bit of its right child ... finally we reach a leaf, whose path indicates the binary representation of $x_1'$. Repeating this procedure, we continue to obtain $x_2', x_3', ..., x_n'$. Hence, we are able to create
the sequence $X'=x_1'x_2'...x_{n-1}'x_n'$ if it exists.

It can be proved that the sequence $X'$ can be successfully constructed if and only the following condition is satisfied: For any $\gamma\in \Upsilon_{b-1}$, $$w_\textrm{T}(X_{\gamma})=|X_{\gamma\textrm{T}}|, \quad w_\textrm{H}(X_{\gamma})=|X_{\gamma\textrm{H}}|,$$
 where $w_\textrm{T}(X)$ counts the number of T's in $X$ and $w_\textrm{H}(X)$ counts the number of H's in $X$.

Obviously, the binary sequences $\{X_{\gamma}\}$ with $\gamma\in \Upsilon_b$ satisfy the above condition. Permuting them into
$\{X_{\gamma}'\}$ with $\gamma\in \Upsilon_b$ does not violate this condition. Hence, we can always construct a sequence
$X'\in \{0,1,..., m-1\}^n$, which yields $\{X_{\gamma}'\}$ with $\gamma\in \Upsilon_b$.

This completes the proof.
\hfill\newQED

Now, we divide all the possible input sequences in $\{0,1,...,m-1\}^n$ into classes. Two sequences $X, X'\in \{0,1,...,m-1\}^n$
are in the same class if and only if the binary sequences obtained from $X$ and $X'$ are permutations with each other, i.e.,
$X_\gamma'$ is a permutation of $X_\gamma$ for all $\gamma\in \Upsilon_b$. Here, we use $\mathbb{G}$ to denote the set consisting of all such classes.

\begin{Lemma} \label{lemma3_2} All the sequences in a class $G\in \mathbb{G}$ have the same probability of being generated.
\end{Lemma}

\IEEEproof Based on the probability distribution of each die roll $\{p_0, p_1, ..., p_{m-1}\}$,
we can get a group of conditional probabilities, denoted as
$$q_{\textrm{T}|\phi}, q_{\textrm{H}|\phi}, q_{\textrm{T}|\textrm{T}}, q_{\textrm{H}|\textrm{T}}, q_{{\textrm{T}|\textrm{H}}}, q_{\textrm{H}|\textrm{H}}, q_{\textrm{T}|\textrm{TT}}, q_{\textrm{H}|\textrm{TT}}, ...,$$
where
$q_{a|\gamma}$ is the conditional probability of generating a die roll $x_i$ such that in its binary representation the bit following a prefix
$\gamma$ is $a$.

Note that $q_{0|\gamma}+q_{1|\gamma}=1$ for all
$\gamma\in \Upsilon_{b}$. For example, if $\{p_0, p_1, p_2\}=\{0.2, 0.3, 0.5\}$,
then
$$q_{0|\phi}=0.5,  q_{0|0}=0.4, q_{0|1}=1.$$

It can be proved that the probability of generating a sequence $X\in\{0,1,..., m-1\}^n$ equals
$$\prod_{\gamma\in \Upsilon_{b}} q_{\textrm{T}|\gamma}^{w_\textrm{T}(X_{\gamma})} q_{\textrm{H}|\gamma}^{w_\textrm{H}(X_\gamma)},$$
where $w_\textrm{T}(X)$ counts the number of T's in $X$ and $w_\textrm{H}(X)$ counts the number of H's in $X$.
This probability keeps unchanged when we permute $X_\gamma$ to $X_\gamma'$ for all $\gamma\in \Upsilon_{b}$.

This implies that all the elements in $G$ have the same probability of being generated.
\hfill\newQED

\begin{Lemma} \label{lemma3_3} Let $\Psi$ be any function that generates random bits from a fixed number of coin tosses.
Given $Z_\gamma, Z_\gamma' \in \{0,1\}^*$ for all $\gamma\in \Upsilon_b$, we define
$$S=\{X| \forall \gamma\in  \Upsilon_b, \Psi(X_\gamma)=Z_\gamma\},$$
$$S'=\{X| \forall \gamma\in  \Upsilon_b, \Psi(X_\gamma)=Z_\gamma'\}.$$
If $|Z_\gamma|=|Z_{\gamma}'|$ for all $\gamma\in \Upsilon_b$, i.e., $Z_\gamma$ and $Z_\gamma'$ have the same length, then for all $G\in \mathbb{G}$,
$$|G\bigcap S|=|G\bigcap S'|,$$
i.e., $G\bigcap S$ and $G\bigcap S'$ have the same size.
\end{Lemma}

\IEEEproof
We prove that for any $\theta\in \Upsilon_b$, if $Z_\gamma=Z_\gamma'$ for all $\gamma\neq\theta$ and
$|Z_\theta|=|Z_{\theta}'|$, then
$$|G\bigcap S|=|G\bigcap S'|.$$
 If this statement is true, we can obtain the
conclusion in the lemma by replacing $Z_\gamma$ with $Z_\gamma'$ one by one for all $\gamma\in \Upsilon_b$.

In the class $G$, assume $|X_\theta|=n_{\theta}$. Let us define $G_{\theta}$
as the subset of $\{0,1\}^{n_{\theta}}$ consisting of all the permutations of $X_{\theta}$.
We also define
$$S_\theta=\{X_{\theta}|\Psi(X_{\theta})=Z_{\theta}\},$$
$$S_\theta'=\{X_{\theta}|\Psi(X_{\theta})=Z_{\theta}'\}.$$
According to Lemma \ref{lemma_coin}, if $\Psi$ can generate random bits from an arbitrary biased coin, then
$$|G_\theta\bigcap S_\theta|=|G_\theta\bigcap S_\theta'|.$$
This implies that all the elements in $G_\theta\bigcap S_\theta$ and those in
$G_\theta\bigcap S_\theta'$ are one-to-one mapping.

Based on this result, we are ready to show that the elements in
$G\bigcap S$ and those in $G\bigcap S'$ are one-to-one mapping:
For any sequence $X$ in $G\bigcap S$, we get a series of binary sequences $\{X_{\gamma}\}$ with $\gamma\in \Upsilon_b$.
Given $Z_{\theta}'$ with $|Z_{\theta}'|=|Z_{\theta}|$, we can find a (one-to-one) mapping of $X_{\theta}$ in $G_\theta\bigcap S_\theta'$, denoted by $X_{\theta}'$. Here, $X_{\theta}'$ is a permutation of $X_\theta$.
According to Lemma \ref{Stream_lemma3_1},
there exists exactly one sequence $X'\in \{0,1,..., m-1\}^n$ such that it yields
$\{X_\phi, X_\textrm{T}, X_\textrm{H},..., X_\theta', ...\}$.  Right now, we see that for any sequence $X$ in
$G\bigcap S$, we can always find its one-to-one mapping $X'$ in $G\bigcap S'$, which implies that
$$|G\bigcap S|=|G\bigcap S'|.$$

This completes the proof.
\hfill\newQED

Based on the lemma above, we get Theorem \ref{theorem3}.

\vspace{0.05in}
\hspace{-0.15in}\textbf{Theorem \ref{theorem3}.} \emph{Let $\Psi$ be any function that generates random bits from a fixed number of coin tosses.
Given a sequence $X\in \{0,1,...,m-1\}^n$ with $m\geq 2$ generated from an $m$-sided die, the generalized scheme of $\Psi$ generates an independent and unbiased sequence.}
\vspace{0.05in}

\IEEEproof
In order to prove that the binary sequence generated is independent and unbiased, we show that for any two sequences $Y_1, Y_2 \in \{0,1\}^k$, they have the same probability to be generated.
Hence, each binary sequence of length $k$ can be generated with probability $\frac{c_k}{2^k}$ for some $0\leq c_k\leq 1$.

First, we let $f:\{0,1,...,m-1\}^n\rightarrow \{0,1\}^*$ be the function of the generalized scheme of $\Psi$, then we write
$$P[f(X)=Y_1]=\sum_{G\in \mathbb{G}}P[f(X)=Y_1,X\in G].$$

According to Lemma \ref{lemma3_2}, all the elements in $G$ have the same probability of being generated. Hence, we denote
this probability as $p_{G}$, and the formula above can written as
$$P[f(X)=Y_1]=\sum_{G\in \mathbb{G}}p_{G}|\{X\in G, f(X)=Y_1\}|.$$

Let $Z_\gamma\in \{0,1\}^*$ be the sequence of bits generated from the node corresponding to $\gamma$ for all $\gamma\in \Upsilon_{b}$,
then $Y_1=\sum_{\gamma\in \Upsilon_{b}} Z_\gamma$. We get that $P[f(X)=Y_1]$ equals
$$\sum_{G\in \mathbb{G}}\sum_{\{Z_\gamma: \gamma\in \Upsilon_{b} \}} p_{G}|\{X\in G,\forall \gamma \in \Upsilon_{b}, \Psi(X_\gamma)=Z_\gamma\}|$$
$$\times I_{\sum_{\gamma\in \Upsilon_{b}} Z_\gamma=Y_1},$$
where $I_{\sum_{\gamma\in \Upsilon_{b}} Z_\gamma=Y_1}=1$ if and only if $\sum_{\gamma\in \Upsilon_{b}} Z_\gamma=Y_1$, otherwise it is zero.

Similarly,
$P[f(X)=Y_2]$ equals
$$\sum_{G\in \mathbb{G}}\sum_{\{Z_\gamma': \gamma\in \Upsilon_{b}\}} p_{G}|\{X\in G,\forall \gamma \in \Upsilon_{b}, \Psi(X_\gamma)=Z_\gamma'\}|$$
$$\times I_{\sum_{\gamma\in \Upsilon_{b}} Z_\gamma'=Y_2},$$

If $|Z_{\gamma}'|=|Z_{\gamma}|$ for all $\gamma\in \Upsilon_b$, then based on Lemma \ref{lemma3_3},
we can get
$$|\{X\in G,\forall \gamma \in \Upsilon_{b}, \Psi(X_\gamma)=Z_\gamma\}|$$
$$=|\{X\in G,\forall \gamma \in \Upsilon_{b}, \Psi(X_\gamma)=Z_\gamma'\}|.$$

Substituting it into the expressions of $P[f(X)=Y_1]$ and $P[f(X)=Y_2]$ shows
$$P[f(X)=Y_1]=P[f(X)=Y_2].$$

So we can conclude that for any binary sequences of the same length, they have the same probability of being generated.
Furthermore, we can conclude that the bits generated are independent and unbiased.

This completes the proof.
\hfill\newQED

\subsection{Optimality}

 In this subsection, we show that the universal scheme keeps the optimality of original algorithms, i.e., if the binary algorithm is asymptotically optimal, like Elias's algorithm or Peres's algorithm, its generalized version is also asymptotically optimal. Here, we say
 an algorithm is asymptotically optimal if and only if the number of random bits generated per input symbol is asymptotically equal to
 the entropy of an input symbol.

\begin{Theorem} \label{Stream_theorem4} Given an $m$-sided die with probability distribution $\rho=(p_0, p_1, ..., p_{m-1})$,
let $n$ be the number of symbols (dice rolls) used in the generalized scheme of $\Psi$ and let $k$ be the number of
random bits generated. If $\Psi$ is asymptotically optimal, then the generalized scheme of $\Psi$ is also asymptotically optimal, that means  $$\lim_{n\rightarrow\infty}\frac{E[k]}{n}=H(\rho),$$
where $$H(\rho)=H(p_0,p_1,...,p_{m-1})=\sum_{i=0}^{m-1}p_i\log_2\frac{1}{p_i}$$
is the entropy of the $m$-sided die.
\end{Theorem}

\IEEEproof We prove this by induction. Using the same notations as above, we have the depth of the binarization tree
$b=\lceil\log_2 m\rceil-1$. If $b=0$, i.e., $m\leq 2$, the algorithm is exactly $\Psi$. Hence, it is asymptotically optimal on efficiency.
Now, assume that the conclusion holds for any integer $b-1$, we show that
it also holds for the integer $b$.

Since the length-$(b+1)$ binary representations of $\{0,1,..., 2^{b}-1\}$ start with $0$,
the probability for a symbol starting with $0$ is
$$ q_0=\sum_{i=0}^{2^{b}-1}p_i.$$
In this case, the conditional probability distribution of these symbols is
$$\{\frac{p_0}{q_0}, \frac{p_1}{q_0}, ..., \frac{p_{2^{b}-1}}{q_0}\}.$$

Similarly, let
$$q_1=\sum_{i=2^{b}}^{m}p_i,$$
then the conditional probability distribution of the symbols starting with $1$ is
$$\{\frac{p_{2^{b}}}{q_1}, \frac{p_{2^{b}+1}}{q_1}, ..., \frac{p_{m-1}}{q_1}\}.$$

When $n$ is large enough, the number of symbols starting with $0$ approaches
$nq_0$ and the number of symbols starting with $1$ approaches $nq_1$. According
to our assumption for $b-1$, the total number of random bits generated approaches
$$nH(q_0, q_1)+ n q_0 H(\frac{p_0}{q_0}, \frac{p_1}{q_0}, ..., \frac{p_{2^{b}-1}}{q_0})$$
$$+nq_1H (\frac{p_{2^{b}}}{q_1}, \frac{p_{2^{b}+1}}{q_1}, ..., \frac{p_{m-1}}{q_1}),$$
which equals
\begin{eqnarray*}
&& n q_0\log_2\frac{1}{q_0}+ n q_1\log_2\frac{1}{q_1} + n q_0 \sum_{i=0}^{2^{b}-1} \frac{p_i}{q_0}\log_2\frac{q_0}{p_i}\\
&&+ n q_1 \sum_{i=2^{b}}^{m-1} \frac{p_i}{q_1}\log_2\frac{q_1}{p_i}\\
&=& n \sum_{i=0}^{m-1}p_i \log_2\frac{1}{p_i}\\
&=& n H(p_0,p_1,...,p_{m-1}).
\end{eqnarray*}

This completes the proof.
\hfill\newQED

\section{Efficient Generation of $k$ Random Bits}
\label{section_coin_kbits}

\subsection{Motivation}

Most existing works on random bits generation from biased coins aim at maximizing the expected number of random bits generated
from a fixed number of coin tosses. Falling into this category, Peres's scheme and Elias's scheme
are asymptotically optimal for generating random bits. However, in these methods, the number of random bits generated is a random variable. In
some occasions, we prefer to generate a prescribed number of random bits, hence it motivates us an opposite question: fixing the number of random bits to generate, i.e., $k$ bits, how can we minimize the expected number of coin tosses? This question is equally important as the original one, since in many applications a prescribed number of random bits are required while the source is usually a stream of coin tosses instead of a sequence of
fixed length. But the existing study on this question is very limited.

To generate $k$ random bits, we are always able to make use of the existing schemes with fixed input length and variable
output length like Peres's scheme or Elias's scheme. For example, we can keep reading $n$ tosses (H or T) for several times and concatenate their outputs until the total number of random bits  generated is slightly larger than $k$. However, if $n$ is small, this approach is
less information efficient. If $n$ is large, this approach may generate too many extra random bits, which can be treated as a waste.
In this section, we propose an algorithm to generate exactly $k$ random bits efficiently. It is motivated by the Elias's scheme.
It can be proved that this algorithm is asymptotically optimal, namely, the expected number of coin tosses required per random bit generated is asymptotically equal to one over the entropy of the biased coin.

\subsection{An Iterative Scheme}
\label{section_Elias}

It is not easy to generate $k$ random bits directly from a biased coin with very high information efficiency.
Our approach of achieving this goal is to generate random bits iteratively -- we first produce
$m\leq k$ random bits, where $m$ is a variable number that is equal to or close to $k$ with very high probability.
In next step, instead of trying to generate $k$ random bits, we try to generate $k-m$ random bits ... we repeat
this procedure until generating total $k$ random bits.

How can we generate $m$ random bits from a biased coin such that $m$ is variable number that is equal to or very close to $k$?
Our idea is to construct a group of disjoint prefix sets, denoted by $S_1,S_2,...,S_w$, such that
(1) all the sequences in a prefix set $S_i$ with $1\leq i\leq w$ have the same probability of being generated, and
(2) $S=S_1\bigcup S_2 \bigcup ...\bigcup S_w$ form a stopping set, namely, we can always get a sequence in $S$ (or with probability almost $1$)
when keeping reading tosses from a biased coin. For example, we can let
\begin{eqnarray*}
  S_1 &=& \{\textrm{HH},\textrm{HT}\}, \\
  S_2 &=& \{\textrm{THH},\textrm{TTT}\}, \\
  S_3 &=& \{\textrm{THT},\textrm{TTH}\}.
\end{eqnarray*}
Then $S=S_1\bigcup S_2 \bigcup S_3$ forms a stopping set, which is complete and prefix-free.

In the scheme, we let all the sequences in $S_i$ for all $1\leq i\leq w$ have the same probability, i.e., $S_i$ consists of
sequences with the same number of H's and T's.  We select criteria carefully such that $|S_i|$ is slightly larger than $2^k$.
Similarly as Elias's original scheme, we assign output binary sequences to all the members in $S_i$ for all $1\leq i\leq w$.
Let $W$ be the number of members that were not assigned yet in a prefix set, then $2^j$ possible output binary sequences of length $j$
are assigned to $2^j$ distinct unassigned members, where $j=k$ if $W\geq 2^k$ and $2^j\leq W<2^{j+1}$ if $W<2^k$.
We repeat the procedure above for the rest of the members
that were not assigned.

\begin{Theorem}
The above method generates $m$ random bits for some $m$ with $0\leq m\leq k$.
\end{Theorem}

\IEEEproof  It is easy to see that the above method never generates a binary sequence longer than $k$. We only need to prove that
for any binary sequences $Y,Y'\in \{0,1\}^m$, they have the same probability of being generated.

Let $f$ denote the function corresponding to the above method. Then
$$P[f(X)=Y]=\sum_{i=1}^wP[X\in S_i] P[f(X)=Y|X\in S_i].$$
Given $X\in S_i$, we have $P[f(X)=Y|X\in S_i]=P[f(X)=Y'|X\in S_i]$, which supports our claim that
any two binary sequences of the same length have the same probability of being generated.
\hfill\newQED

The next question is how to construct such prefix sets $S_1, S_2, ..., S_w$. Let us first consider the construction
of their union, i.e., the stopping set $S$. Given a biased coin, we design an algorithm that reads coin tosses and stops the reading until
it meets the first input sequence that satisfies some criterion. For instance, let $k_1$ be the number of $\textrm{H}$'s and $k_2$ be the number of $\textrm{T}$'s in the current input sequence, one possible choice is to read coin tosses until we get the first sequence such that $\nchoosek{k_1+k_2}{k_1}\geq 2^k$.
Such an input sequence is a member in the stopping set $S$. However, this criterion is not the best one that we can have, since it will introduce too many iterations to generate $k$ random bits. To reduce the number of iterations, we hope that the size of each prefix set, saying $S_i$, is slightly larger than $2^k$. As a result, we use the following stopping set:
$$S=\{\textrm{the first sequence s.t. } \nchoosek{k_1+k_2}{k_1}\geq \frac{2^k(k_1+k_2)}{\min(k_1,k_2)}\}.$$
Later, we will show that the selection of such a stopping set can make the number of iterations very small.

Now we divide all the sequences in the stopping set $S$ into different classes, i.e., the prefix sets $S_1, S_2, ..., S_w$, such that
each prefix set consists of the sequences with the same number of H's and T's.  Assume $S_{k_1,k_2}$ is a nonempty prefix set that
consists of sequences with $k_1$ H's and $k_2$ T's, then
$$S_{k_1,k_2}= G_{k_1,k_2} \bigcap S,$$
where $G_{k_1,k_2}$ is the set consisting of all the sequences with $k_1$ H's and $k_2$ T's. According to the stopping set constructed above, we
have
$$S_{k_1,k_2}= \{x\in G_{k_1,k_2}|  \nchoosek{k_1+k_2}{k_1}\geq \frac{2^k(k_1+k_2)}{\min(k_1,k_2)}, $$
$$\nchoosek{k_1+k_2-1}{k_1'}<\frac{2^k(k_1+k_2-1)}{\min(k_1',k_2')} \},$$
where $k_1'$ is the number of $\textrm{H}$'s in $x$ without considering the last symbol and $k_2'$
is the number of $\textrm{H}$'s in $x$ without considering the last symbol. So if the
last symbol of $x$ is $\textrm{H}$, then $k_1'=k_1-1, k_2'=k_2$; if
the last symbol of $x$ is $\textrm{T}$, then $k_1'=k_1, k_2'=k_2-1$.
 According to the expression of $S_{k_1,k_2}$,
we see that the sequences in a prefix set are not prefixes of sequences in another prefix set. Furthermore, we can prove that the size of each prefix
set is at least $2^k$.

\begin{Lemma}\label{bias_lemma_1}
If $S_{k_1,k_2}\neq \phi$, then $|S_{k_1,k_2}|\geq 2^k$.
\end{Lemma}

\IEEEproof Without loss of  generality, we assume that $k_1\leq k_2$, hence, $\nchoosek{k_1+k_2}{k_1} \geq \frac{2^k(k_1+k_2)}{k_1}$. It also implies $k_1\geq 1$.
To prove  $|S_{k_1,k_2}|\geq 2^k$, we show that $S_{k_1,k_2}$ includes
all the sequences $x\in G_{k_1,k_2}$ ending with  $\textrm{H}$.
If $x\in G_{k_1,k_2}$ ending with $\textrm{H}$ does not belong to $S_{k_1,k_2}$, then
$$\nchoosek{k_1+k_2-1}{k_1'}\geq \frac{2^k(k_1+k_2-1)}{k_1'}.$$ From which, we can get
$$\nchoosek{k_1+k_2-1}{k_1}\geq \frac{2^k(k_1+k_2-1)}{k_1}.$$ It further implies that
all the sequences $x\in G_{k_1,k_2}$ ending with  $\textrm{T}$ are also not members in $S_{k_1,k_2}$. So
$S_{k_1,k_2}$ is empty. It is a contradiction.

The number of sequences $x\in G_{k_1,k_2}$ ending with  $\textrm{H}$ is
$$\nchoosek{k_1+k_2-1}{k_1-1}=\nchoosek{k_1+k_2}{k_1}\frac{k_1}{k_1+k_2}\geq 2^k.$$
So the size of $S_{k_1,k_2}$ is at least $2^k$ if $S_{k_1,k_2}\neq \phi$.
This completes the proof.
\hfill\newQED

Based on the construction of prefix sets, we can get an algorithm $\Phi_{k}$ for generating $m$ random bits with $0\leq m\leq k$, described as follows.

\vspace{0.05in}
\begin{list}{\labelitemi}{\leftmargin=0.5em}
\renewcommand{\labelitemi}{}
\item \textbf{Algorithm $\mathbf{\Phi_k}$}
\item \textbf{Input:} A stream of biased coin tosses.
\item \textbf{Output:} $m$ bits with $0\leq m\leq k$.

\begin{algorithmic}
\STATE \textbf{(1)} Reading coin tosses until there are $k_1$ H's and $k_2$ T's for some $k_1$ and $k_2$ such that
$$\nchoosek{k_1+k_2}{k_1}\geq \frac{2^k(k_1+k_2)}{\min(k_1,k_2)}.$$
\STATE \textbf{(2)} Let $X$ denote the current input sequence of coin tosses. If the last coin toss is $\textrm{H}$, we let $k_1'=k_1-1, k_2'=k_2$; otherwise,
we let $k_1'=k_1, k_2'=k_2-1$. We remove this coin toss from $X$ if
$$\nchoosek{k_1+k_2-1}{k_1'}\geq \frac{2^k(k_1+k_2-1)}{\min(k_1',k_2')}.$$
\STATE \textbf{(3)} Let $\Psi_E$ denote the Elias's function\footnote{Here, an arbitrary algorithm for generating random bits from a fixed number of coin tosses works.} for generating random bits from a fixed number of coin tosses. A fast computation
of $\Psi_E$ was provided by Ryabko and Matchikina in \cite{Ryabko2000}. The output of the algorithm $\Psi_k$
is $\Psi_E(X)$ or the last $k$ bits of $\Psi_E(X)$ if $\Psi_E(X)$ is longer than $k$.
\end{algorithmic}
\end{list}
\vspace{0.05in}

According to Lemma \ref{bias_lemma_1}, we can easily get the following conclusion.
\begin{Corollary} \label{bias_corollary_2}
The algorithm $\Phi_k$ generates $m$ random bits for some $m$ with $0\leq m\leq k$, and $m=k$ with probability at least $1/2$.
\end{Corollary}

\IEEEproof The sequence generated by $\Phi_k$ is independent and unbiased. This conclusion is immediate from Lemma \ref{bias_lemma_1}.
Assume that the input sequence $x\in S_i$ for some $i$ with $1\leq i\leq w$, then the probability of $m=k$ is
$$\frac{\lfloor\frac{|S_i|}{2^k}\rfloor 2^k}{|S_i|},$$
which is at least $1/2$ based on the fact that $|S_i|\geq2^k$.
Since this conclusion is true for all $S_i$ with $1\leq i\leq w$, we can claim that $m=k$ with probability at least $1/2$.
\hfill\newQED

Since the algorithm $\Phi_k$ generates $m$ random bits for some $m$ with $0\leq m\leq k$ from an arbitrary biased coin, we
are able to generate $k$ bits iteratively: After generating $m$ random bits, we apply the algorithm $\Phi_{k-m}$ for generating $k-m$ bits. Repeating this procedure, the total number of random bits generated will converge to $k$ very quickly. We call this scheme as an iterative scheme for generating $k$ random bits.

To generate $k$ random bits, we do not want to iterate $\Phi_k$ too many times. Fortunately, in the following theorem,
we show that in our scheme the expected number of iterations is upper bounded by a constant $2$.

\begin{Theorem}
The expected number of iterations in the iterative scheme for generating $k$ random bits is at most $2$.
\end{Theorem}

\IEEEproof According to Corollary \ref{bias_corollary_2}, $\Phi_k$ generates $m=k$ random bits with probability at least $1/2$.
Hence, the scheme stops at each iteration with probability more than $1/2$.
Following this fact, the result in the theorem is immediate.
\hfill\newQED

\subsection{Optimality}

In this subsection, we study the information efficiency of the iterative scheme and show that this scheme is asymptotically optimal.

\begin{Lemma} Given a biased coin with probability $p$ being $\textrm{H}$, let $n$ be the number of coin tosses used by the algorithm $\Phi_k$, then
$$\lim_{k\rightarrow \infty} \frac{E[n]}{k}\leq \frac{1}{H(p)}.$$
\end{Lemma}

\IEEEproof We consider the probability of having an input sequence of length at least $n$, denote as $P_n$.
In this case, we can write $n=k_1+k_2$, where $k_1$ is the number of H's and $k_2$ is the number of T's.
According to the construction of the stopping set,
$${\nchoosek{n-1}{\min(k_1,k_2)-1}}<2^k\frac{n-1}{\min(k_1,k_2)-1}.$$
Or we can write it as
$${\nchoosek{n-2}{\min(k_1,k_2)-2}}<2^k.$$

Hence, we get an upper bound for $\min(k_1,k_2)$, which is
\begin{equation}\label{equ_lemma1_1}
t_n=\max\{i\in\{0,1,...,n\}| {\nchoosek{n-2}{i-2}}<2^k\}.
\end{equation}

Note that if ${\nchoosek{n-2}{\frac{n}{2}-2}}\geq 2^k$, then $t_n$ is a nondecreasing function of $n$.

According to the symmetry of our criteria, we can get
$$P_n \leq \sum_{i=0}^{t_n} (p^i(1-p)^{n-i}+(1-p)^i p^{n-i}) {\nchoosek{n}{i}}.$$

For convenience, we write
$$Q_n= \sum_{i=0}^{t_n} (p^i(1-p)^{n-i}+(1-p)^i p^{n-i}) {\nchoosek{n}{i}},$$
then $P_n\leq Q_n$ and $Q_n$ is also a nondecreasing function of $n$.

Now, we are ready to calculate the expected number of coin tosses required, which equals
\begin{eqnarray}
E[n]
&=&\sum_{n=1}^{\infty} (P_n-P_{n+1}) n=\sum_{n=1}^{\infty}P_n\label{equ_lemma1_5}\\
&\leq & \sum_{n=1}^{\frac{k}{H(p)}(1+\epsilon)}P_n+ \sum_{n=\frac{k}{H(p)}(1+\epsilon)}^{\infty}Q_n+\sum_{n=2\frac{k}{H(p)}(1+\epsilon)}^\infty Q_n, \nonumber
\end{eqnarray}
where $\epsilon>0$ is a small constant. In the rest, we study the upper bounds for all the three terms when $n$ is large enough.

For the first term, we have
\begin{equation} \label{equ_lemma1_2}
\sum_{n=1}^{\frac{k}{H(p)}(1+\epsilon)}P_n\leq \frac{k}{H(p)}(1+\epsilon).
\end{equation}

Now let us consider the second term
$$\sum_{n=\frac{k}{H(p)}(1+\epsilon)}^{2\frac{k}{H(p)}(1+\epsilon)}Q_n \leq \frac{k}{H(p)}(1+\epsilon) Q_{\frac{k}{H(p)}(1+\epsilon)}.$$

Using the Stirling bounds on factorials yields
$$ \lim_{n\rightarrow \infty }\frac{1}{n}\log_2 {\nchoosek{n}{\rho n}}=H(\rho),$$
where $H$ is the binary entropy function. Hence, following (\ref{equ_lemma1_1}), we can get
$$ \lim_{n\rightarrow \infty} H(\frac{t_n}{n})=\lim_{n\rightarrow\infty}\frac{k}{n}.$$

When $n=\frac{k}{H(p)}(1+\epsilon)$, we can write
$$\lim_{n\rightarrow \infty} H(\frac{t_n}{n})=\frac{H(p)}{1+\epsilon},$$
which implies that
\begin{equation*}
\lim_{n\rightarrow\infty}\frac{t_n}{n}=p-\epsilon_1
\end{equation*}
for some $\epsilon_1>0$. So there exists an $N_1$ such that for $n>N_1$,  $\frac{n_t}{n}\leq p-\epsilon_1/2$.

By the weak law for the binomial distribution, given any $\epsilon_2>0$ and $\delta>0$, there
is an $N_2$ such that for $n >N_2$, with probability at least $1-\delta$ there are $i$ H's among the $n$ coin tosses
such that $|\frac{i}{n}-p|\leq \epsilon_2$. Letting $\epsilon_2=\epsilon_1/2$ and $n=\frac{k}{H(p)}(1+\epsilon)$ gives
$$Q_n\leq \delta,$$
for any $\delta>0$ when $n>\max(N_1,N_2)$.

So for any $\delta>0$, when $k$ is large enough, we have
\begin{equation} \label{equ_lemma1_3}
\sum_{n=\frac{k}{H(p)}(1+\epsilon)}^{2\frac{k}{H(p)}(1+\epsilon)}Q_n \leq \frac{k}{H(p)}(1+\epsilon) \delta.
\end{equation}

To calculate the third term, we notice that $Q_n$ decays very quickly as $n$ increase when $n\geq 2\frac{k}{H(p)}(1+\epsilon)$.
In this case,
\begin{eqnarray*}
&&\frac{Q_{n+1}}{Q_n}\\
&=&\frac{\sum_{i=0}^{t_{n+1}} (p^i(1-p)^{n+1-i}+(1-p)^i p^{n+1-i}) {\nchoosek{n+1}{i}}}{\sum_{i=0}^{t_n} (p^i(1-p)^{n-i}+(1-p)^i p^{n-i}) {\nchoosek{n}{i}}}\\
&\leq& \frac{\sum_{i=0}^{t_{n}} (p^i(1-p)^{n+1-i}+(1-p)^i p^{n+1-i}) {\nchoosek{n+1}{i}}}{\sum_{i=0}^{t_n} (p^i(1-p)^{n-i}+(1-p)^i p^{n-i}) {\nchoosek{n}{i}}}\\
&\leq & \max_{i=0}^{t_n} \frac{(p^i(1-p)^{n+1-i}+(1-p)^i p^{n+1-i}) {\nchoosek{n+1}{i}}}{(p^i(1-p)^{n-i}+(1-p)^i p^{n-i}) {\nchoosek{n}{i}}}\\
&\leq &(1-p) \max_{i=1}^{t_n}\frac{n+1}{n+1-t_n}\\
&\leq & \frac{(1-p)n}{n-t_n}.
\end{eqnarray*}

When $n\geq 2\frac{k}{H(p)}(1+\epsilon)$, we have
$$\lim_{n\rightarrow\infty} H(\frac{t_n}{n})=\lim_{n\rightarrow\infty}\frac{k}{n}\leq \frac{H(p)}{2(1+\epsilon)}.$$
This implies that when $n$ is large enough, $H(\frac{t_n}{n})\leq \frac{H(p)}{2}$. Let us define a constant $\alpha$ such that $\alpha\leq \frac{1}{2}$ and $H(\alpha)=\frac{H(p)}{2}$. Then for all $n\geq 2\frac{k}{H(p)}(1+\epsilon)$, when $k$ is large enough,
$$\frac{Q_{n+1}}{Q_n}\leq \frac{1-p}{1-\alpha}<1.$$

Therefore, given any $\delta>0$, when $k$ is large enough, the value of the third term
\begin{eqnarray}
\sum_{n=2\frac{k}{H(p)}(1+\epsilon)}^\infty Q_n
&\leq & Q_{2\frac{k}{H(p)}(1+\epsilon)} \sum_{i=0}^\infty (\frac{1-p}{1-\alpha})^i \nonumber\\
&\leq & Q_{\frac{k}{H(p)}(1+\epsilon)} \frac{1}{1-\frac{1-p}{1-\alpha}} \nonumber\\
&\leq & \frac{1-\alpha}{p-\alpha}\delta. \label{equ_lemma1_4}
\end{eqnarray}

Substituting (\ref{equ_lemma1_2}), (\ref{equ_lemma1_3}), and (\ref{equ_lemma1_4}) into (\ref{equ_lemma1_5}) yields that for any $\epsilon>0$ and $\delta>0$,
if $k$ is large enough, we have
$$E[n]\leq \frac{k}{H(p)}(1+\epsilon)(1+\delta)+ \frac{1-\alpha}{p-\alpha}\delta,$$
with $\alpha<p$.

Then it is easy to get that
$$ \lim_{k\rightarrow\infty}\frac{E[n]}{k}\leq \frac{1}{H(p)}.$$
This completes the proof.
\hfill\newQED

\begin{Theorem} Given a biased coin with probability $p$ being H, let $n$ be the number of coin tosses required to generate $k$ random bits in the iterative scheme, then
$$\lim_{k\rightarrow\infty}\frac{E[n]}{k}= \frac{1}{H(p)}.$$
\end{Theorem}

\IEEEproof First, we prove that $\lim_{k\rightarrow\infty}\frac{E[n]}{k}\geq \frac{1}{H(p)}$.
Let $X\in\{0,1\}^*$ be the input sequence, then
$$\lim_{k\rightarrow\infty}\frac{E[n]H(p)}{H(X)}=1.$$

Shannon's theory tells us that it is impossible to extract more than $H(X)$ random bits from $X$, i.e., $H(X)\geq k$.
So $$\lim_{k\rightarrow\infty} \frac{E[n]}{k}\geq \frac{1}{H(p)}.$$

To get the conclusion in the theorem, we only need to show that
$$\lim_{k\rightarrow\infty}\frac{E[n]}{k}\leq\frac{1}{H(p)}.$$

To distinguish the $n$ in this theorem and the one in the previous theorem, we use $n_{(k)}$ denote the number of coin tosses required to
generate $k$ random bits in the iterative scheme and let $n_{(k)}^\Phi$ denote the number of coin tosses required by
$\Phi_k$. Let $p_m$ be the probability for $\Phi_k$ generating $m$ random bits with $0\leq m\leq k$. Then we have that
\begin{equation} \label{equ_lemma2_1}
E[n_{(k)}]=E[n_{(k)}^{\Phi}]+\sum_{m=0}^k p_m E[n_{(k-m)}].
\end{equation}

According to the algorithm, $p_k\geq \frac{1}{2}$ and $E[n_{(k-m)}]\leq E[n_{(k)}]$. Substituting them into the equation above gives
$$E[n_{(k)}]\leq E[n_{(k)}^{\Phi}]+\frac{1}{2} E[n_{(k)}],$$
i.e., $E[n_{(k)}]\leq 2 E[n_{(k)}^{\Phi}]$.

Now, we divide the second term in (\ref{equ_lemma2_1}) into two parts such that
$$E[n_{(k)}]\leq E[n_{(k)}^{\Phi}]+\sum_{m=0}^{k-\epsilon k} p_m E[n_{(k-m)}]+ \sum_{m=k-\epsilon k}^k p_m E[n_{(k-m)}],$$
for a constant $\epsilon>0$. In which,
$$\sum_{m=0}^{k-\epsilon k} p_m E[n_{(k-m)}]\leq (\sum_{m=0}^{k-\epsilon k} p_m)2 E[n_{(k)}^{\Phi}],$$
$$\sum_{m=k-\epsilon k}^k p_m E[n_{(k-m)}]\leq 2E[n_{(\epsilon k)}^{\Phi}].$$

Hence
\begin{equation}\label{equ_lemma2_4}
E[n_{(k)}]\leq E[n_{(k)}^{\Phi}]+  (\sum_{m=0}^{k-\epsilon k} p_m)2 E[n_{(k)}^{\Phi}] + 2E[n_{(\epsilon k)}^{\Phi}].
\end{equation}

Given $k$, all the possible input sequences are divided into $w$ prefix sets $S_1, S_2, ..., S_w$,
where $w$ can be an infinite number. Given an input sequence $X\in S_i$ for $1\leq i\leq w$, we are considering
the probability for $\Phi_k$ generating a sequence of length $m$.

In our algorithm, $|S_i|\geq 2^k$. Assume
$$|S_i|=\alpha_k 2^k+ \alpha_{k-1} 2^{k-1}+...+\alpha_0 2^0,$$
where $\alpha_k\geq 1$ and $0\leq \alpha_0, \alpha_1, ...,\alpha_{k-1}\leq 1$. Given the condition $X\in S_i$,
we have
$$\sum_{m=0}^{k-\epsilon k} p_m= \frac{\sum_{i=0}^{k-\epsilon k} \alpha_i 2^ i}{\sum_{i=0}^k \alpha_i 2^ i}\leq \frac{2^{k-\epsilon k+1}}{2^k+2^{k-\epsilon k+1}}\leq \frac{2^{k-\epsilon k+1}}{2^k}.$$
So given any $\delta>0$, when $k$ is large enough, we have
\begin{equation}\label{equ_lemma2_2}
\sum_{m=0}^{k-\epsilon k} p_m\leq \delta.
\end{equation}
Although we reach this conclusion for $X\in S_i$, this conclusion holds for any $S_i$ with $0\leq i\leq w$. Hence, we are able to remove this
constrain that $X\in S_i$.

According to the previous lemma, for any $\delta>0$, when $k$ is large enough, we have
\begin{equation}\label{equ_lemma2_3}
\frac{E[n_{(\epsilon k)}^{\Phi}]}{\epsilon k}\leq \frac{1}{H(p)}+\delta,
\end{equation}
\begin{equation}\label{equ_lemma2_5}
\frac{E[n_{( k)}^{\Phi}]}{k}\leq \frac{1}{H(p)}+\delta.
\end{equation}

Substituting (\ref{equ_lemma2_2}), (\ref{equ_lemma2_3}), and (\ref{equ_lemma2_5}) into (\ref{equ_lemma2_4}) gives us
$$E[n_{(k)}]\leq k(\frac{1}{H(p)}+\delta)(1+2\delta) +2 k\epsilon(\frac{1}{H(p)}+\delta).$$
From which, we obtain
$$\lim_{k\rightarrow \infty}\frac{E[n]}{k}=\lim_{k\rightarrow \infty}\frac{E[n_{(k)}]}{k}\leq \frac{1}{H(p)}.$$
This completes the proof.
\hfill\newQED

The theorem above shows that the iterative scheme is asymptotically optimal, i.e., the expected number of
coin tosses for generating $k$ random bits approaches the information theoretic bound by below when $k$ becomes large.

\section{Conclusion}
\label{section_coin_conclusion}

In this paper, we have presented a universal scheme that transforms an arbitrary algorithm for $2$-faced coins to generate random bits from general $m$-sided dice,
hence enabling the application of existing algorithms to general sources. Although a similar question has been studied before, as in \cite{Juels2000}, their solution can only be applied to a specified algorithm, i.e., Elias's algorithm.

The second contribution of this paper is an efficient algorithm for generating a prescribed number of random bits from an arbitrary biased coin.
In many applications, this is a natural way of considering the problem of random bits generation from biased coins, but it is not well studied in the literature.
This problem is similar to the one studied in universal variable-to-fixed length codes, which are used to parse an infinite sequence into variable-length phases.
Each phase is then encoded into a fixed number of bits. In \cite{Lawrence1977}, Lawrence devised a variable-to-fixed length code for the class of binary memoryless sources (biased coins),
which is based on Pascal's triangle (so is our algorithm). Tjalkens and Willems \cite{Tjalkens1992} modified Lawrence's algorithm as a more natural and simple implementation, and they showed that the rate of the resulting code converges asymptotically optimally fast to the source entropy. These universal variable-to-fixed length codes are probably capable to generate random bits asymptotically in some (week) sense, namely,
the random bits generated in this way are not perfect, and they cannot satisfy the typical requirement based on statistical distance (widely used in computer science).

%
%
%
%
%




\end{document}